# Intense boundary emission destroys normal radio-frequency plasma sheath

Guang-Yu Sun, An-Bang Sun, and Guan-Jun Zhang[*]

*Research Center for Advanced High Voltage and Plasma Technology, State Key Laboratory of Electrical Insulation and Power Equipment, Xi'an Jiaotong University, Xi'an, Shaanxi 710049, China*

Plasma sheath is the non-neutral space charge region that isolates bulk plasma from boundary. Radio-frequency (RF) sheathes are formed when applying RF voltage to electrodes. Generally, applied bias is mainly consumed by RF sheath which shields external field. Here we report first evidence that intense boundary emission destroys normal RF sheath and establishes a novel type of RF plasma where external bias is consumed by bulk plasma instead of sheath. Ions are naturally confined while plasma electrons are unobstructed, generating strong RF current in entire plasma, combined with unique particle and energy balance. Proposed model offers possibility for ion erosion mitigation of plasma-facing component. It also inspires technics for reaction rate control in plasma processing and wave mode conversion.

## 1. Introduction

Plasma sheath is one of the most ubiquitous and well-known features that permeates laboratory plasmas as most confined plasmas have boundaries[1-5]. Dating back to around a century ago when Irving Langmuir first introduced the term "sheath" to describe an ion-rich region adjacent to the electrode[6], studies of sheath have since been developing rapidly and are still of vital importance in modern plasma physics. Sheath matters in fusion devices[1, 2, 7, 8], dusty plasma[3, 9, 10], spacecraft propulsion[11, 12], plasma diagnostics, material processing[13, 14] and many others[5, 15, 16].

Applying radio-frequency (RF) voltage to electrodes creates capacitively coupled plasma (CCP) with two oscillating RF sheathes. Understanding RF sheath is of fundamental interest and is essential in numerous applications[13, 14, 17-21]. Contemporary models of CCP assume that the bulk plasma is well isolated from boundaries by sheathes, and applied bias mainly rests on sheathes instead of bulk plasma. Conduction current dominates over displacement current in bulk plasma while field in bulk plasma is shielded. However, in this work we will show that the bias can be primarily consumed by bulk plasma and electric field in plasma center needs not to be shielded by sheathes, due to intense boundary emission.

Multitudes of studies have been performed regarding boundary emission in CCP, but its influences are mostly assumed to be unessential[20, 22-29]. The steady flux of ion $\varGamma_i$ produces a surface emission flux $\varGamma_{em} = \gamma_i \varGamma_i$ due to ion-induced secondary electron emission (SEE), with $\gamma_i$ the ion-induced secondary electron emission coefficient. Under low pressure, secondary electrons (SEs) are lost before remarkable ionizations occur and their impact is less significant than that of hot electrons generated by stochastic heating[30]. Higher pressure incurs transition to γ mode where ionizations of SEs become dominant[24]. Yet the structure of plasma, i.e. bulk plasma connected by Bohm presheath and Child-law sheath, is retained and mean wall potential remains negative relative to plasma[31-35].

---

[*]Email address: gjzhang@xjtu.edu.cn

In this work, we show that intense boundary emission incurs remarkable influences and restructures entire RF plasma, which happens when $\Gamma_{em}$ is greater than plasma electron flux $\Gamma_{ep}$ in average. It is well known that a space-charge limited (SCL) sheath is formed if $\Gamma_{em} > \Gamma_{ep}$ near plasma boundary, such as thermionic emitter, emissive probe, Hall thruster, etc[36-38]. In this case, normal Child-law sheath still exists between plasma and the local potential minimum called virtual cathode (VC), so plasma dynamics are not essentially modified[39]. Recent works reported that a floating SCL sheath cannot remain stable due to cold ion accumulation in VC[40, 41]. But no one has studied a RF plasma with intense boundary emission. Below we shall first show with simulation how RF plasma behaves under strong boundary emission, a theoretical ground is then given to validate simulation results. Model generality, practical methods for implementation, prospects in future applications are discussed as well.

## 2. Model set-up

To investigate RF plasma with boundary emission, we employ a 1D1V continuum kinetic simulation code which advances kinetic equation and solves Poisson equation in each step. The use of continuum, kinetic simulation has been proved powerful in numerous works related to boundary plasma physics.[42-46] Kinetic simulation produces smooth data free from noise and allows convenient control of plasma density, temperature and velocity distribution function (VDF), etc. In our simulation, RF plasma bounded by two parallel planar electrodes is considered. Sinusoidal source is fixed as Dirichlet-type condition at two boundaries. Initial state is uniform Maxwellian distribution in space for both electrons and ions, i.e. $f_{s0}(x,v) = n_0 \sqrt{\frac{m_s}{2\pi T_s}} \exp(-\frac{m_s v_s^2}{2T_s})$. Here subscript $s$ represents ion ($i$) or electron ($e$), $n_0$ is initial plasma density, $m, T, v$ are mass, temperature, velocity, respectively. After simulation begins, the VDFs are advanced according to the kinetic equation below:

$$\frac{\partial f_s(x,v)}{\partial t} + v \frac{\partial f_s(x,v)}{\partial x} + \frac{q_s[E+v\times B]}{m_s} \frac{\partial f_s(x,v)}{\partial v} = S_{charge} + S_{collision,s} \qquad (2.1)$$

with $q_s$ the particle charge, $E$ and $B$ fields at location $x$, $f_s$ the distribution function depending on time $t$, location $x$ and velocity $v$. Equation 2.1 includes 4 advections which dictate the evolution of VDF, namely $-v\frac{\partial f_s(x,v)}{\partial x}$, $-\frac{q_s[E(x)+v\times B]}{m_s}\frac{\partial f_s(x,v)}{\partial v}$, $S_{charge}$ and $S_{collision,s}$. They are space advection, velocity advection, source term and collision term, respectively. Note that the time step, space and velocity resolution are controlled to satisfy Courant–Friedrichs–Lewy stability condition. Also, the velocity advection requires the solution of 1D Poisson equation, which solved by finite difference method. Detailed definitions of source and collision terms are provided below.

Source term compensates particle loss at boundaries. One approach is to generate electron-ion pairs according to velocity-dependent ionization, i.e. $S_{charge} = f_{s0}v_e n_g \sigma_{iz}$ with $v_e$ electron velocity (more precisely relative velocity), $n_g$ background gas density and $\sigma_{iz}$ cross section determined by electron kinetic energy. In our simulation, the plasma density is expected to be controlled as constant to facilitate comparison with theory, so the following expression is adopted:

$$S_{charge} = \frac{\Gamma_{i,L}+\Gamma_{i,R}}{n_0 L_s} f_{s0} \tag{2.2}$$

with $\Gamma_{i,L/R}$ ion flux at left/right wall and $L_s$ the length where source is supplied. In this way, ion lost at boundaries is compensated by Maxwellian electron-ion pairs generated by source term.

Collision term characterizes the change of distribution function due to particle collision. Collision term for electrons is defined as follows:

$$S_{collision,e} = \frac{v_{Tep}}{\lambda_e}\left(\frac{n_e}{n_0}f_{e0} - f_e\right) \tag{2.3}$$

where $v_{Tep} = \sqrt{\frac{T_{ep}}{m_e}}$, $T_{ep}$ is temperature of plasma electrons, $\lambda_e$ is electron mean free path and $n_e$ is electron density at position $x$. With Eq. 2.3, electrons that encounter collision are replaced by equal number of electrons following Maxwellian distribution. $\lambda_e$ is set to be greater than sheath size and much smaller than total length of simulation domain, representing typical collisionless sheath. For ions the following operator is used:

$$S_{collision,i} = \frac{\int_{-\infty}^{+\infty} v_i f_i dv_i}{n_0}\frac{f_{i0}}{\lambda_i} - \frac{|v_i|}{\lambda_i}f_i \tag{2.4}$$

where $\lambda_i$ is ion mean free path usually greater than sheath size, $v_i$ is ion velocity. Equation 2.4 represents charge exchange collisions for ions, which removes fast ions from $f_i$ at each position $x$ and replaces them with an equal number of cold ions following $f_{i0}$. The charge exchange collision frequency is proportional to ion velocity. Validity of above two collision terms has been justified in previous works.[40, 47, 48]

Regarding surface emission, it is characterized by boundary condition of EVDF in the following way (taking left boundary as example):

$$f_e(x=0, v_e>0) = \frac{m_e}{T_{em}}\Gamma_{em}\exp\left(-\frac{m_e v_e^2}{2T_{em}}\right) \tag{2.5}$$

Here $\Gamma_{em}$ is surface emission flux and $T_{em}$ is temperature of emitted electrons. For thermionic emission and photoemission, $\Gamma_{em}$ is a constant depending on electrode temperature or light intensity. For ion distribution, RHS of Equation 2.5 is zero. Regarding secondary electron emission induced by electrons, it is:

$$f_e(x=0, v_e > 0) = \frac{m_e}{T_{em}} \gamma_e \Gamma_{ep} \exp(-\frac{m_e v_e^2}{2T_{em}}) \tag{2.6}$$

where $\gamma_e = \frac{\Gamma_{em}}{\Gamma_{ep}}$ electron-induced secondary electron emission coefficient.

Detailed adopted parameters are given below to help replicate presented data. $n_0 = 5\times10^{14}$ m$^{-3}$, $T_{ep} = 14$ eV, $T_{em} = 2.5$ eV, $T_i = 0.05$ eV, ion mass 1u. $\lambda_e = 1.25$ cm and $\lambda_i = 0.3$ cm. $L = 10$ cm, space resolution $\Delta x = 10^{-4}$ m. Range of electron/ion velocity is $4v_{Tep}$ and $2c_s$, with $v_{Tep} = \sqrt{\frac{T_{ep}}{m_e}}$ and $c_s = \sqrt{\frac{T_{ep}}{m_i}}$, divided into 601 points. Time step is $\Delta t = 2\times10^{-3}$ ns, source frequency is 13.56 MHz.

## 3. Results

For normal RF sheath, the flux balance in average should be fulfilled, which means that $\langle \Gamma_{ep} - \Gamma_i - \Gamma_{em} \rangle_T = 0$. In typical RF discharge, $\omega_i \ll \omega \ll \omega_e$ with $\omega_i, \omega_e$ ion and electron frequency, and $\omega$ the source frequency. In this case, ions respond to time-averaged space potential so flux balance is simplified as $\langle \Gamma_{ep} - \Gamma_{em} \rangle_T = \Gamma_i = n_{sh} u_B$ where $n_{sh}$ is plasma density at sheath edge and $u_B$ is Bohm velocity. Bohm velocity is crucial in plasma sheath theory. Its value is influenced by collisionality, while a recent study of Tang and Guo indicates that electron heat flux also plays a critical role.[49]

However, when boundary emission is greater than plasma electron flux, the normal flux balance no longer holds since ion flux should be nonnegative, which indicates that classical CCP breaks down when $\langle \Gamma_{ep} - \Gamma_{em} \rangle_T < 0$. In order to investigate the RF structure under intense boundary emission, simulation is performed with $\langle \Gamma_{ep} - \Gamma_{em} \rangle_T < 0$, and a constant $\Gamma_{em} = 10^{21} m^{-2} s^{-1}$ is used (roughly 2 times $\Gamma_{ep}$). Results are shown Fig.1a where two tiny sheath barriers with positive potential relative to sheath edge present near the surfaces, and nearly entire applied bias is absorbed by bulk plasma. We call this new RF plasma *inverted RF plasma* (IRP) as its sheath potential is inverted relative to normal CCP. Note that IRP is fundamentally different from previously observed field reversal in CCP, as the latter is a local and transient effect so the global CCP properties remain unchanged[50, 51].

Simulation shows that electric field in bulk plasma equals to $E_{trans} = \frac{V_{RF}}{d} \cos(\omega t)$ with $d$ the half of gap distance and $V_{RF}$ half of total applied voltage amplitude. In Fig. 1b total current of normal CCP leads that of inverted RF plasma by nearly $\frac{\pi}{2}$, the latter coincides with source voltage in phase. This phenomenon is highly unusual as RF sheath in CCP contains capacitance $C_{sh} = a\frac{\varepsilon_0 A}{s_{sh}}$ with $A$ the electrode area, $\varepsilon_0$ vacuum permittivity, $s_{sh}$ sheath size and $a$ a constant depending on model assumptions[30]. In normal CCP the voltage across two capacitive sheaths far exceeds plasma voltage, and current is $\frac{\pi}{2}$ ahead of voltage. Field is much stronger in sheath where strong displacement current $J_d$ fulfills the continuity. Note that in bulk of CCP the conduction current $J_c$ is dominant as only weak field penetrates into bulk plasma. Fig. 1c shows a mode transition from CCP to IRP by increasing $\Gamma_{em}$. If the sheath potential in normal CCP is regarded as positive, one can find that the sign of sheath potential is inverted when $\Gamma_{em}$ gradually exceeds $\Gamma_{ep}$, which is marked as mode transition in Fig. 1c. The sign of wall electric field is inverted as well.

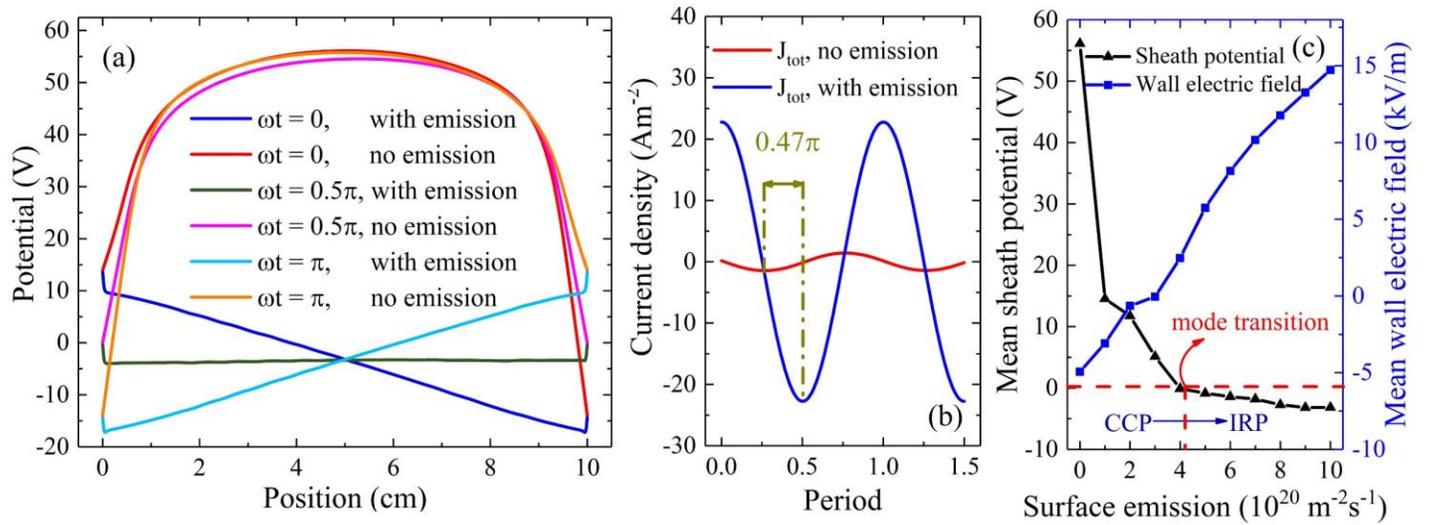

Figure 1. Simulation results of RF plasma with/without boundary emission. In (a) potentials of normal CCP are compared to condition with surface emission of $10^{21}$ m$^{-2}$s$^{-1}$, which is around twice the initial plasma electron flux. Applied bias is consumed by bulk plasma and two small sheaths opposite to normal RF sheathes appear at both ends. In (b) the total currents of RF discharge with/without strong boundary emission are compared. They have a phase difference of nearly $0.5\pi$ and boundary emission enlarges total current by more than 1 order of magnitude. External source is $V_{RF}\cos(\omega t)$. Panel (c) shows the transition from normal RF plasma (CCP) to inverted RF plasma when increasing $\Gamma_{em}$. Mean sheath potential and mean electric field of left wall are recorded with different boundary emissions. For convention, mean sheath potential in CCP is chosen as positive, which becomes negative after $\langle \Gamma_{ep} - \Gamma_{em}\rangle_T < 0$ (marked as mode transition). Sign of wall electric field is also inverted after mode transition.

The reason why an IRP behaves as above can be understood considering flux of plasma electrons. When wall potential is positive relative to sheath edge, plasma electron flux is unobstructed. Conduction current and displacement current in normal CCP and IRP are shown in Fig. 2a, b for comparison. Wall potential is negative relative to sheath edge in CCP, repelling most plasma electrons back hence the current continuity in sheath must be fulfilled by strong displacement current. This is, however, unnecessary when wall potential is positive. Intense conduction current travels unimpeded in bulk plasma as well as in sheath, which is calculated by $J_c = \sigma_p E_{trans} = \frac{n_0 e^2}{m_e \overline{v_m}} \frac{V_{RF}}{d}\cos(\omega t) \approx 27.87\cos(\omega t)$ A/m$^2$. Here $e$ is elementary charge, $\sigma_p$ is plasma conductivity, $E_{trans}$ is bulk electric field amplitude and $\overline{v_m}$ is collision frequency. Amplitude given by simulation is 23.02 A/m$^2$, which is not far from theory.

Flux balance in IRP becomes clear with above analyses. Fig. 2c, d give the distribution functions of two types of RF plasma. Clearly, presheath and sheath accelerate ions and repel plasma electrons in CCP. Conversely, in IRP no Bohm presheath presents and mean sheath potential is positive relative to sheath edge, hence ions are confined but plasma electrons are unimpeded. Meanwhile, $\Gamma_i$ is diminished in sheath while $\Gamma_{ep}$ conserves. Emitted electrons from boundary are partially reflected back to surface by inverse sheath barrier and some penetrate into plasma, leading to intense electron concentration near boundary. The flux balance is therefore written as $\langle \Gamma_{ep} + \Gamma_{eref} - \Gamma_{em} \rangle_T = 0$ with $\Gamma_{eref}$ the flux of reflected emitted electrons towards boundary. $\Gamma_{em}$ is constant in simulation, possibly representing thermionic emission or photoemission, but can also be configured as a function of $\Gamma_{ep}$ due to SEE[52]. The unique particle and power balance in IRP will be analyzed following a theoretical ground to be established later on.

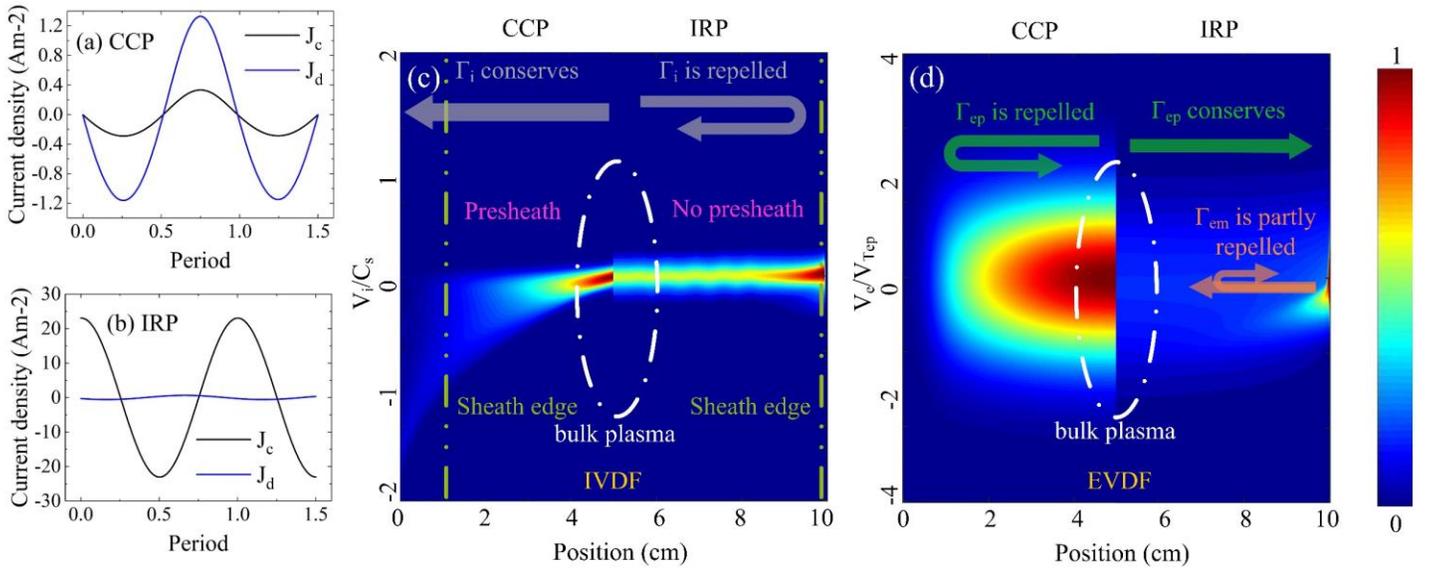

Figure 2. (a), (b) show $J_c$ and $J_d$ in sheath of CCP and IRP. $J_d \gg J_c$ in RF sheath of CCP while $J_c \gg J_d$ in inverted RF sheath. (c)(d) give normalized ion and electron velocity distribution functions when $\omega t = 0.5\pi$ for CCP (left panel) and IRP (right panel). Velocity space is normalized to $v_{Tep} = \sqrt{\frac{T_{ep}}{m_e}}$ and $c_s = \sqrt{\frac{T_{ep}}{m_i}}$. In (c) left panel shows that ions are accelerated towards boundary in CCP and a presheath exists which ensures a minimum velocity when ions enter the sheath. Right panel shows that there is no presheath in IRP and ions are repelled back to plasma by inverse sheath barrier. The size of RF sheath in CCP is greater than in IRP. In (d) left panel shows that plasma electrons are repelled in CCP while right panel shows that in IRP plasma electrons are unimpeded and part of emitted electrons are repelled back to surface, forming an intense electron concentration near surface.

To summarize, when average plasma electron flux is below boundary emission ($\langle \Gamma_{ep} - \Gamma_{em} \rangle_T < 0$), normal CCP is replaced by IRP because conventional flux balance ($\langle \Gamma_{ep} - \Gamma_i - \Gamma_{em} \rangle_T = 0$) cannot be fulfilled. IRP contains two oscillating electron-rich inverse sheaths where wall potential relative to sheath edge is positive. Consequently, ions are confined while plasma electrons are unimpeded. Meanwhile, the lack of sheath capacitance as in CCP is combined with much stronger conduction current which dominates in both bulk plasma and sheath. Below a theoretical model will be constructed based on obtained simulation results.

Step function model is widely used in many CCP theories, which assumes a time-dependent sheath edge out of which electron density equals to zero. In such model a Child-law type sheath and Bohm presheath are chosen as prerequisite[32, 33, 53, 54]. However, existing CCP theories are no longer valid for IRP and new theoretical ground should be established. Below we will first deduce inverse sheath potential and then solve the whole IRP. Two key parameters are defined as follows: (1) mean inverse sheath $\overline{\varphi_{inv}}$ representing mean potential of plasma center relative to wall; (2) temporal inverse sheath $\varphi_{inv}(\omega t)$ representing real-time wall potential relative to sheath edge. The former characterizes ion dynamics and the latter is for electrons. Both parameters are chosen as positive for simplicity. In Fig. 3 definitions of adopted parameters are shown.

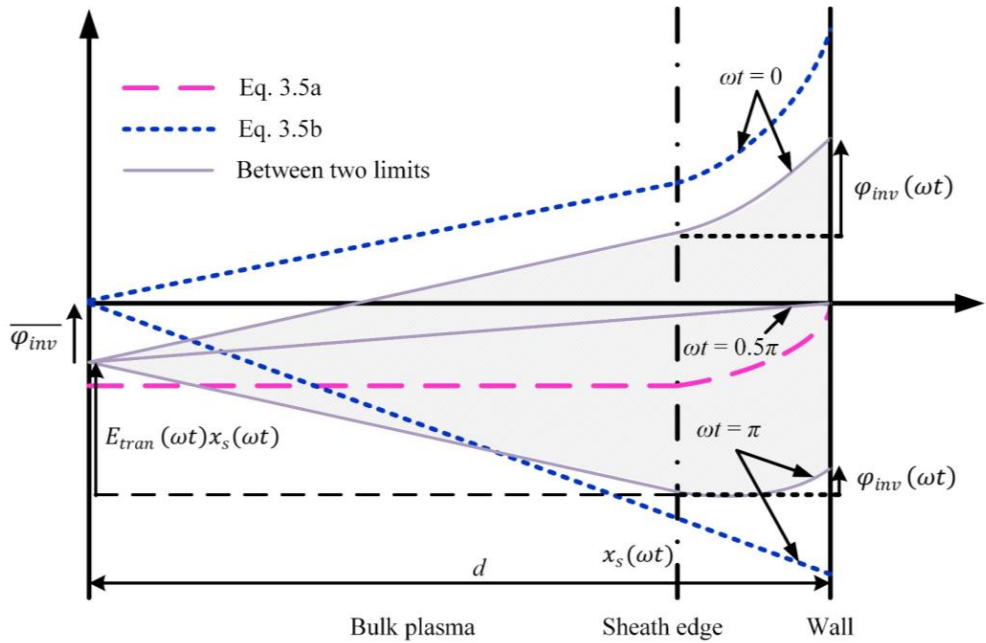

Figure 3. Schematic of adopted theoretical model for inverted RF plasma. $\overline{\varphi_{inv}}$ is mean potential of plasma center relative to wall, $\varphi_{inv}(\omega t)$ is real-time wall potential relative to sheath edge, and $E_{tran}(\omega t)x_s(\omega t)$ is sheath edge potential relative to plasma center. Two limiting cases dictated by Eq. 3.5 are plotted plus normal situations between two limits. Dashed pink curve represents limit of Eq. 3.5a when no external source is given, which is reduced to a floating inverse sheath. Two dotted blue curves represent the limit of Eq. 3.5b with infinite external source, where mean inverse sheath barrier is eliminated and bulk plasma is directly linked with boundary without sheath in half period. Note that the black dash dotted line is only an approximation of inverse sheath edge, actually the position of sheath edge $x_s(\omega t)$ is time-dependent, the same as in CCP.

Considering real-time boundary condition $V_{wall} = \tilde{V}\cos(\omega t)$, potential difference between plasma center and boundary is expressed as follows according to Fig. 3:

$$-\overline{\varphi_{inv}} + E_{tran}(\omega t)x_s(\omega t) + \varphi_{inv}(\omega t) = V_{RF}\cos(\omega t) \qquad (3.1)$$

with $x_s$ the location of sheath edge, so sheath size is $d - x_s$, $d$ is the half of gap distance. A simplified expression of temporal inverse sheath barrier is obtained according to Poisson equation:

$$\varphi_{inv}(\omega t) = \frac{\beta e n_{sh}}{2\varepsilon_0}[d - x_s(\omega t)]^2 \qquad (3.2)$$

where $\beta$ is an adjustable parameter generally equal or greater than 1, representing electron density $\beta n_{sh}$ in inverted RF sheath. Note that $n_{sh}$ is electron density at sheath edge. Ions are assumed to be cold and are not considered in inverse sheath. The assumption used in Eq. 3.2 is analogous to a previous inverse sheath model for floating boundary.[55]

Using basic plasma kinetic theories, densities of plasma electrons $n_{ep}$, emitted electrons $n_{em}$, and reflected emitted electrons $n_{eref}$ in inverse sheath are derived as follows. Note that for convention, we choose $\varphi = 0$ at sheath edge, though sheath edge potential is time-dependent relative to the mean potential of boundary.

$$n_{ep}(\varphi) = n_{ep0}\exp(\frac{e\varphi}{T_{ep}})\text{erfc}(\sqrt{\frac{e\varphi}{T_{ep}}}) \tag{3.3a}$$

$$n_{em}(\varphi) = n_{emw}\exp[\frac{e(\varphi-\varphi_{inv}(\omega t))}{T_{em}}] \tag{3.3b}$$

$$n_{eref}(\varphi) = n_{em}(\varphi)\text{erf}(\sqrt{\frac{e\varphi}{T_{em}}}) \tag{3.3c}$$

Here $n_{ep0}$ is plasma electron density at sheath edge and $n_{emw}$ is emitted electron density at wall, deductions of Eq. 3.4 are based on integration of EVDF from corresponding velocity range which can be commonly found in many related works of sheath physics[37, 55-57]. Taking both emitted electron and plasma electron flux as half-Maxwellian, we obtain $\Gamma_{ep} = n_{ep0}\sqrt{\frac{2T_{ep}}{\pi m_e}}$, $\Gamma_{em} = n_{emw}\sqrt{\frac{2T_{em}}{\pi m_e}}$. Reflected electron flux is $\Gamma_{eref} = \Gamma_{em}[1 - \exp(-\frac{e\varphi_{inv}(\omega t)}{T_{em}})]$, which is the complement of emitted electrons penetrating the temporal inverse sheath $\Gamma_{em}\exp(-\frac{e\varphi_{inv}(\omega t)}{T_{em}})$.

In addition, the charge neutrality at sheath edge must hold, which gives $n_{ep}(\varphi) + n_{em}(\varphi) + n_{eref}(\varphi)\big|_{sheath\ edge} = n_{sh}$. Combining Eq. 3.1-3.3 and charge neutrality, the following relation is obtained:

$$\varphi_{inv}(\omega t) = \frac{(M+\sqrt{M^2+4\overline{\varphi_{inv}}})^2}{4} \tag{3.4}$$

with $M = \sqrt{\frac{2\varepsilon_0}{\beta e n_{sh}}}\frac{V_{RF}}{d}\cos(\omega t)$. It is then possible to calculate both $\varphi_{inv}(\omega t)$ and $\overline{\varphi_{inv}}$ with aforementioned flux balance in average $\frac{1}{2\pi}\int_0^{2\pi}(\Gamma_{ep} + \Gamma_{eref} - \Gamma_{em})d(\omega t) = 0$. Validity of above deductions can be briefly justified by dimension and the following two limits:

$$\overline{\varphi_{inv}}|_{V_{RF}=0} = \varphi_{inv}(\omega t)|_{V_{RF}=0} = \frac{T_{em}}{e}\ln(\frac{\Gamma_{em}}{\Gamma_{ep}}) \tag{3.5a}$$

$$\varphi_{inv}(\omega t)|_{V_{RF}\to+\infty} = \begin{cases} M^2, \omega t \in [k\pi, (k+0.5)\pi], k \in Z^{\geq} \\ 0, \omega t \notin [k\pi, (k+0.5)\pi], k \in Z^{\geq} \end{cases} \tag{3.5b}$$

Limits in Eq. 3.5 show that the RF sheath is reduced to floating inverse sheath when $V_{RF} = 0$. Temporal inverse sheath is rectified in half period when $V_{RF} \to +\infty$. Once $\overline{\varphi_{inv}}$ and $\varphi_{inv}(\omega t)$ are determined, potential in entire IRP can be calculated by solving Poisson's equation, which requires an order reduction with the numerical integral below:

$$\left(\frac{d\varphi}{dx}\right)^2 \Big|_0^\varphi = \frac{2n_{ep0}T_{ep}}{\varepsilon_0}\left[\exp\left(\frac{e\varphi}{T_{ep}}\right) - 1 - \mathcal{F}(\varphi, T_{ep})\right] + \frac{2n_{emw}T_{em}}{\varepsilon_0}\exp\left(-\frac{e\varphi_{inv}(\omega t)}{T_{em}}\right)\left[\exp\left(\frac{e\varphi}{T_{em}}\right) - 1 + \mathcal{F}(\varphi, T_{em})\right] \quad (3.6)$$

where $n_{emw}$ is given by $\Gamma_{em}$, $n_{ep0}$ is solved from charge neutrality and $\mathcal{F}(\varphi, T) = \exp\left(\frac{e\varphi}{T}\right)\text{erf}\left(\sqrt{\frac{e\varphi}{T}}\right) - 2\sqrt{\frac{e\varphi}{\pi T}}$. The integral in Eq. 3.6 is solved from $\frac{d^2\varphi}{dx^2} = \frac{e}{\varepsilon_0}[n_{ep}(\varphi) + n_{em}(\varphi) + n_{eref}(\varphi)]$ multiplied by $\frac{d\varphi}{dx}$ then integrating over $x$. Calculated potentials are given in Fig. 3a, showing good agreement with simulation. The discrepancy of $\overline{\varphi_{inv}}$ is due to assumptions of ion-free sheath and collisionless sheath, which are not exactly the case in simulation. Fig. 3b shows $\varphi_{inv}(\omega t)$ with different source voltages. They are normalized by $\overline{\varphi_{inv}}$ to facilitate comparison. For small $V_{RF}$ it is collinear with sinusoidal source but it gradually approaches the limit of Eq. 3.5b and becomes rectified at half period when $V_{RF}$ is large. A complete rectification occurs when $V_{RF} \to +\infty$, yet higher ionization rate may shift the realistic condition from the ideal limit. Calculated $\overline{\varphi_{inv}}$ is shown in Fig. 3c. The mean sheath barrier rises up with both $\Gamma_{em}$ and $V_{RF}$, making it possible to control the ion confinement by changing boundary emission and source amplitude. Note that the emission threshold above which IRP is formed increases with $V_{RF}$.

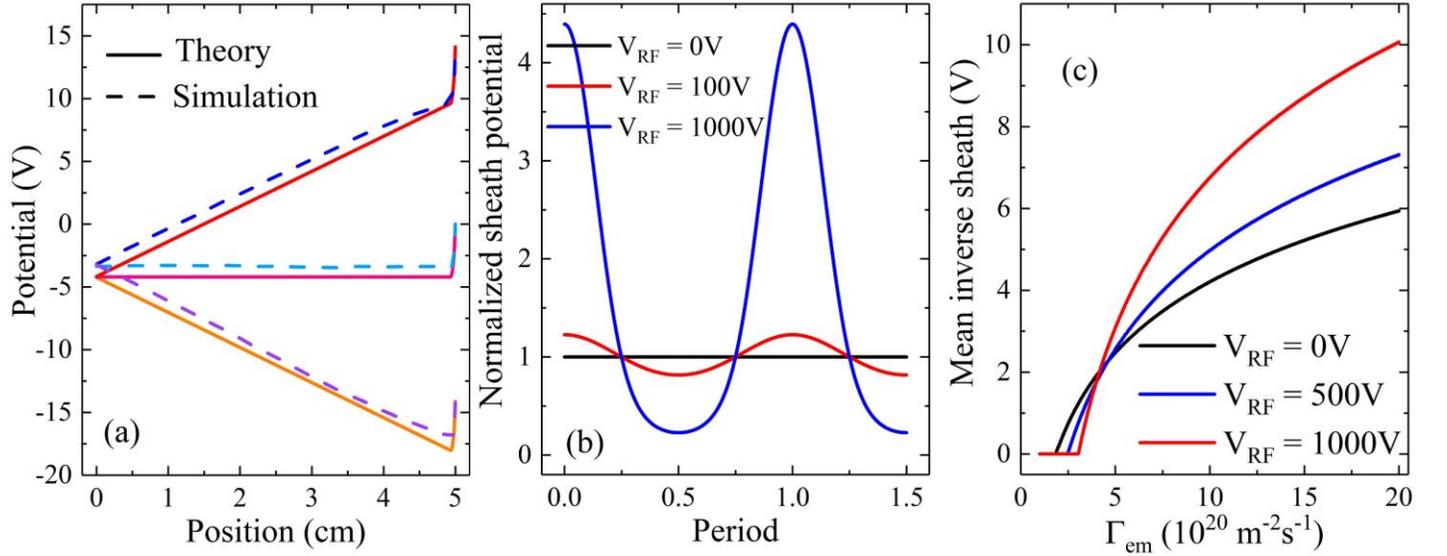

Figure 4. Results given by theory. In (a) the same parameters as in simulation are used to calculated space potential in IRP at $\omega t = 0, 0.5\pi, \pi$. (b) gives $\frac{\varphi_{inv}(\omega t)}{\overline{\varphi_{inv}}}$ at different $V_{RF}$, unit V. $V_{RF} = 0$ gives floating inverse sheath (Eq. 3.5a), sheath potential is sinusoidal for small $V_{RF}$, for large $V_{RF}$ it is rectified in half period (Eq. 3.5b). (c) shows $\overline{\varphi_{inv}}$ with different $\Gamma_{em}$ and $V_{RF}$. The minimum $\Gamma_{em}$ to invoke IRP increases with $V_{RF}$, and $\overline{\varphi_{inv}}$ increases with $\Gamma_{em}$.

Abovementioned sheath structure of IRP leads to unique particle and power balance with respect to normal CCP. For particle balance, ionizations should compensate for ion loss at boundaries. Ion flux conserves in CCP while only energetic ions crossing $\overline{\varphi_{inv}}$ can hit the wall in IRP, which makes particle balance different between two types of RF plasma. General expression of particle balance is $n_0 K_{iz} n_g d_b = 2\Gamma_{iw}$, with $K_{iz}$ the ionization rate coefficient and $\Gamma_{iw}$ the ion flux at wall. Here $d_b = 2\langle x_s \rangle_T$ is the region where ionization occurs, which excludes two sheaths. Note that $\langle x_s \rangle_T$ is the mean RF sheath location relative to plasma center. In CCP, $\Gamma_{iw} = n_{sh} u_B$ and $n_{sh} = h_l n_0$ with $h_l$ the sheath edge to center density ratio. Note that here the IVDF does not play a role because $T_i \ll e\overline{\varphi_{CCP}}$ with $\overline{\varphi_{CCP}}$ the mean RF sheath potential in CCP (positive). In IRP, it is better to involve IVDF since if not, no ion can pass the mean inverse sheath barrier $\overline{\varphi_{inv}}$ if $T_i < e\overline{\varphi_{inv}}$, which is hardly realistic if ion temperature becomes remarkable (though not exceeding $e\overline{\varphi_{inv}}$). We consider IVDF in inverse sheath edge as follows:

$$f_i(v_i) = n_{sh} \begin{cases} \dfrac{\sqrt{\frac{2m_i}{\pi T_i}}}{1+\text{erf}(\sqrt{\frac{e\overline{\varphi_{inv}}}{T_i}})} \exp\left(-\frac{m_i v_i^2}{2T_i}\right), & v_i \geq v_{i,min} \\ 0, & v_i < v_{i,min} \end{cases} \tag{3.7}$$

where $v_{i,min} = -\sqrt{\frac{2\overline{\varphi_{inv}}}{m_i}}$. The fact that $f_i = 0$ when $v_i < v_{i,min}$ is due to ion loss cone, i.e. energetic ions which cross inverse sheath barrier cannot return, while those less energetic ions are reflected back. Ion flux towards wall at sheath edge is therefore expressed as $\Gamma_{i0} = \dfrac{n_{sh}\sqrt{\frac{2T_i}{\pi m_i}}}{1+\text{erf}(\sqrt{\frac{e\overline{\varphi_{inv}}}{T_i}})}$. Accordingly, ion wall flux in IRP is $\Gamma_{iw} = \Gamma_{i0} \exp\left(-\frac{e\overline{\varphi_{inv}}}{T_i}\right) = \dfrac{n_{sh}\sqrt{\frac{2T_i}{\pi m_i}}}{1+\text{erf}(\sqrt{\frac{e\overline{\varphi_{inv}}}{T_i}})} \exp\left(-\frac{e\overline{\varphi_{inv}}}{T_i}\right)$. Apparently, since ion wall flux in IRP is smaller than in CCP, IRP can be sustained with weaker ionization. This also indicates that some ions inevitably flow to boundary if $T_i$ is remarkable. In deductions of IRP structure it is assumed that no ion presents in inverse sheath, which is based on the assumption that $T_i \ll T_{em}$. It has been shown that higher ion temperature slightly increases $\overline{\varphi_{inv}}$.[55]

The power balance is the balance between power absorption and power loss. The former consists ohmic heating $S_{ohm}$ and stochastic heating $S_{stoc}$ in CCP, while the latter is due to particle loss at boundary $S_{edge}$ and interparticle collisions $S_{coll}$. Electron power balance in CCP is written as:[22]

$$S_{coll} + S_{edge} = S_{ohm} + S_{stoc} = S_e \tag{3.8}$$

indicating power lost due to collisions and boundary flux is equal to power gained from ohmic heating and stochastic heating. Note that $S$ terms are power per unit area of electrode. $S_e$ is the power gain/loss of electrons. The LHS is frequently combined as $2\Gamma_{ew} e(\mathcal{E}_c + \mathcal{E}_{e,loss})$ where $\Gamma_{ew}$ is electron flux at wall, $\mathcal{E}_c$ is collisional energy loss per created electron–ion pair and $\mathcal{E}_{e,loss}$ is the kinetic energy lost per electron lost at boundary. $\mathcal{E}_c$ is defined by relation below:[30]

$$K_{iz}\mathcal{E}_c = K_{iz}\mathcal{E}_{iz} + K_{ex}\mathcal{E}_{ex} + K_{el}\frac{2m_e}{m_i}T_{ep} \qquad (3.9)$$

Subscript $iz$, $ex$ and $el$ represent ionization, excitation, and elastic scattering against neutral atoms, with $K$ and $\mathcal{E}$ the rate coefficients and energy losses of different collision types. In CCP $\mathcal{E}_{e,loss} = 2T_{ep}$ and corresponding ion energy loss $\mathcal{E}_{i,loss} = 2T_i + \frac{T_{ep}}{2} + e\overline{\varphi_{CCP}}$ where $2T_i$ is frequently neglected as it is much smaller than the others. For IRP, the expression of $\mathcal{E}_c$ is unchanged. For plasma electrons $\mathcal{E}_{ep,loss} = e\overline{\varphi_{inv}} + 2T_{ep}$ and for ion $\mathcal{E}_{i,loss} = 2T_i$. (Rigorously, $\mathcal{E}_{ep,loss}(\omega t)$ is time-dependent because electrons respond to instantaneous potential variation, here we take the average value for simplicity, same as follows) This is because the wall potential relative to sheath edge in IRP is inverted with respect to RF sheath in CCP, hence incident energies of plasma electrons and ions are inverted as well.

Regarding power absorption, ohmic heating is similar in both types of RF discharge, but IRP contains higher ohmic heating power as its conduction current is intense. It is worthwhile to mention that the stochastic heating is zero in inverted RF plasma since its presheath potential is flat. Stochastic heating in normal CCP discharge is due to non-synchronous motion of sheath edge and bulk plasma, which can be verified by current continuity $n_{sh}v_{e,sheath\ edge} = n_0 v_{e,bulk}$, with $v_{e,sheath\ edge}$, $v_{e,bulk}$ electron velocity at sheath edge and bulk plasma, $n_{sh}$, $n_0$ density at sheath edge and bulk plasma, respectively.[58] The drop of plasma density in Bohm presheath makes $v_{e,sheath\ edge} \neq v_{e,bulk}$ so a velocity modulation takes place, which cannot happen in inverted RF plasma where presheath potential is flat and plasma electron density is uniform. Aforementioned expressions between two types of RF plasma are summarized in Table I for comparison.

A major difference between CCP without boundary emission and IRP is that in IRP there exist a power gain from surface emission and an additional power loss from flux of reflected surface emission. The power balance of electron therefore has to be modified as follows:

$$S_{coll} + S_{edge} = S_{ohm} + S_{emis} = S_e \qquad (3.10)$$

where $S_{emis}$ is the power of surface emission and is expressed as $S_{emis} = 2\Gamma_{em}e\mathcal{E}_{e,em}$ with $\mathcal{E}_{e,em}$ the kinetic energy gain per emitted electron from boundary. Note that $S_{edge}$ should contain both plasma electrons and reflected emitted electrons, i.e. $S_{edge} = S_{edge,ep} + S_{edge,eref}$. Here $S_{edge,ep} = 2\Gamma_{epw}e\mathcal{E}_{ep,loss}$ and $S_{edge,em} = 2\Gamma_{eref}e\mathcal{E}_{eref,loss}$ with $\Gamma_{epw}$ plasma electron wall flux and $\mathcal{E}_{eref,loss}$ the kinetic energy lost per reflected electron at boundary. The expression of $\mathcal{E}_{e,em}$ and $\mathcal{E}_{eref,loss}$ depends on EVDF of emitted electrons. For half-Maxwellian emitted electrons, $\mathcal{E}_{e,em} = \frac{T_{em}}{2}$ and $\mathcal{E}_{eref,loss} = \langle \frac{1}{\text{erf}\left(\sqrt{\frac{e\varphi_{inv}(\omega t)}{T_{em}}}\right)}[\frac{T_{em}}{2} - \sqrt{\frac{e\varphi_{inv}(\omega t)T_{em}}{\pi}}\exp(-\frac{e\varphi_{inv}(\omega t)}{T_{em}})]\rangle_T$. The expression of $\mathcal{E}_{eref,loss}$ is calculated from distribution function $f_{eref}(v_e) = n_{eref} \frac{\sqrt{\frac{2m_e}{\pi T_{em}}}}{\text{erf}(\sqrt{\frac{e\varphi_{inv}(\omega t)}{T_{em}}})} \exp\left(-\frac{m_e v_e^2}{2T_{em}}\right)$ for $0 < v_e \leq \sqrt{\frac{2e\varphi_{inv}(\omega t)}{m_e}}$ and 0 otherwise (configuration is the same as in Fig. 3 with bulk plasma on the left and wall on the right). The missing high velocity tail in $f_{eref}$ comes from the fact that emitted electrons which cross the inverse sheath cannot return. To complete the discussion on power balance, we note that the total RF power is $S_{RF} = S_e + 2\Gamma_{iw} e \mathcal{E}_{i,loss}$ with the second term on RHS power loss due to ion wall flux. This indicates that total absorbed RF power get lost through ion wall flux and electron power dissipation. The latter can be further divided into boundary flux loss and collisional power loss. The distinct sheath structure in IRP makes its power balance quite different from CCP.

TABLE I. Ion wall flux, kinetic energy lost per plasma electron lost at boundary, kinetic energy lost per ion lost at boundary, ohmic heating power, stochastic heating power in CCP and IRP. $\overline{\varphi_{CCP}}$ is mean sheath potential in CCP (positive), $d_b = 2\bar{x}_s$ is the region where ionization occurs (excluding two sheaths).

| Type | $\Gamma_{iw}$ | $\mathcal{E}_{ep,loss}$ | $\mathcal{E}_{i,loss}$ | $S_{ohm}$ | $S_{stoc}$ |
|---|---|---|---|---|---|
| CCP | $n_{sh} u_B$ | $2T_{ep}$ | $2T_i + \frac{T_{ep}}{2} + e\overline{\varphi_{CCP}}$ | $\frac{1}{2}J^2 \frac{d_b}{\sigma_p}$ | nonzero |
| IRP | $\frac{n_{sh}\sqrt{\frac{2T_i}{\pi m_i}}}{1 + \text{erf}(\sqrt{\frac{e\overline{\varphi_{inv}}}{T_i}})} \exp\left(-\frac{e\overline{\varphi_{inv}}}{T_i}\right)$ | $e\overline{\varphi_{inv}} + 2T_{ep}$ | $2T_i$ | $\frac{1}{2}J^2 \frac{d_b}{\sigma_p}$ | 0 |

## 4. Discussions

An interesting topic still under debate is about the sheath solution for strongly emissive surface ($\Gamma_{em} > \Gamma_{ep}$). Though in this work we report the formation of inverse sheath in RF plasma, another sheath solution, i.e. SCL sheath, which contains non-monotonic space potential also exists in numerous studies of plasma-surface interaction. Typically, a potential well is formed in front of electron-emitting surface to reflect part of emission back. Bohm criterion is still valid in this condition. Actually the sheath is similar to classical sheath if counted between potential minimum and bulk plasma. It is seen near moon and spacecraft surface due to secondary electron emission or photoemission.[59, 60] Similar results are also observed in positively-charged dust grain where conventional orbital-motion-limited (OML) theory is no longer valid,[10, 61] which can be revised by a OML+ theory.[10] Various potential shapes near macroparticle changing with ion angular moment are also reported by Keider et al.[62] In addition, SCL sheath due to thermionic emission is reported in emissive probe.[37]

In order to control the transition between CCP and IRP, a first method could be monitoring the surface emission, which is considerably difficult if emission is triggered by secondary electron emission or thermionic emission. The former requires a sudden change of electron energy while the latter demands a rapid variation of electrode temperature. The fact that there exist two plasma equilibrium states when boundary emission exceeds plasma electron flux brings us a possibility to perform mode transition easily. It has been shown that under floating or DC condition both SCL and inverse sheath could exist under certain circumstance.[45, 48] A SCL sheath can be maintained only if no cold ion accumulates within the virtual cathode (the potential well), elsewise VC will be filled due to cold ion trapping and flux balance will be broken. The cold ion generation is usually caused by charge exchange collision or ionization, hence one may expect a SCL sheath in RF plasma if the ion collision and source terms are closed.

Simulation results show that above speculations are correct, which implies that IRP mode can be switched on and off flexibly, as shown in Fig. 5. Simulation in Fig. 5 begins with initial $\Gamma_{em} = 5\times10^{19} m^{-2} s^{-1}$ ($\langle\Gamma_{ep} - \Gamma_{em}\rangle_T > 0$). After discharge becomes stable, $\Gamma_{em}$ is suddenly augmented to $5\times10^{20} m^{-2} s^{-1}$ ($\langle\Gamma_{ep} - \Gamma_{em}\rangle_T < 0$) to trigger the mode transition. In Fig. 5a the classical RF sheath gradually collapses after transition starts. A stable inverted RF plasma is achieved after decades of periods. In simulation of Fig. 5b electron-ion pairs are not generated in boundary adjacency, also the charge exchange collision is forced to close. It is clear that the SCL RF sheath is formed and bulk plasma is well isolated from the wall similar to normal CCP. To realize the switch on/off of inverted RF plasma, one can make use of E×B drift or applying perpendicular electric field, such that created cold ions can escape from the potential well in other dimensions.

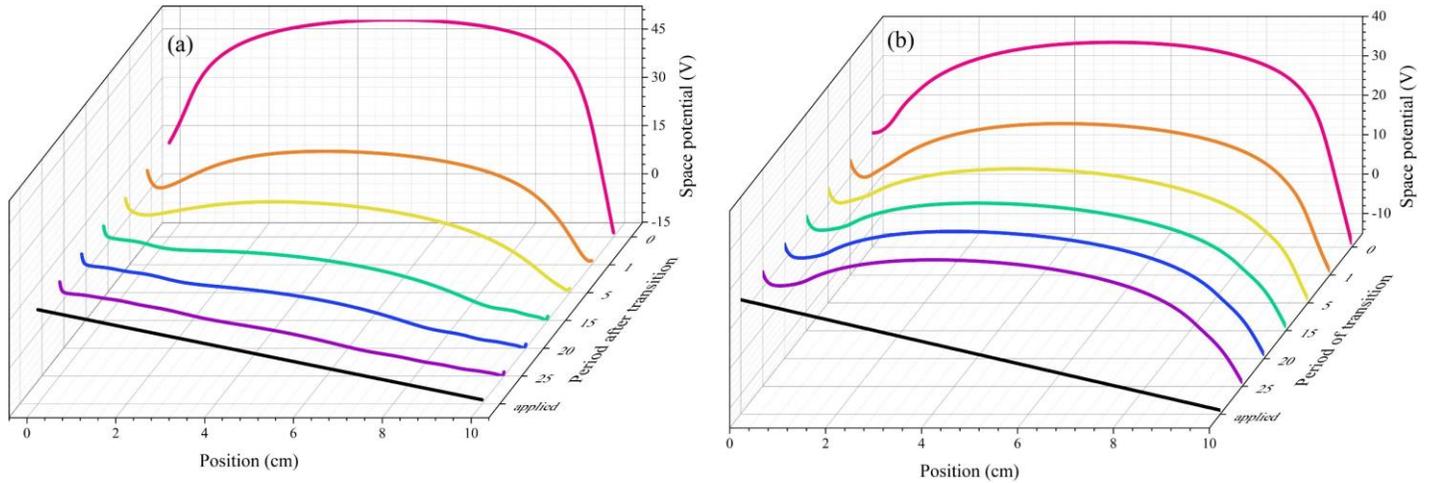

Figure 5. Time evolution of mode transition. Initially $\Gamma_{em} = 5\times10^{19} m^{-2} s^{-1}$ is given until a stable normal RF sheath is formed. $\Gamma_{em}$ is suddenly changed to $\Gamma_{em} = 5\times10^{20} m^{-2} s^{-1}$ to initiate the transition. Space potentials within decades of periods during transition are recorded, straight black curve (front) is applied potential profile in vacuum serving as reference. All data are taken at $\omega t = 0$. (a) Inverted RF plasma mode is switched on, no presheath exists in steady state, inverse sheath barriers are clearly seen. The bulk plasma is poorly shielded by sheaths and applied field is unimpeded in plasma. (b) Inverted RF plasma mode is switched off, RF sheath is space-charge limited, ions are unconfined, Bohm criterion is valid and bulk plasma is well isolated from the boundary with weak transiting field in plasma bulk.

To sustain an IRP in practice, one can capitalize on boundary emission though SEE[63-65], thermionic emission[40, 45, 46] and photoemission[66-68]. In boundaries of many RF-heated plasma systems, ion flux is damaging, e.g. plasma thruster[69], tokamak edge region[70-72], etc. Ion flux induces wall erosion and impurity influx[73, 74], which may be eradicated by invoking IRP. The fact that $\overline{\varphi_{inv}}$ rises up with applied voltage makes it possible to confine hot ions with limited boundary emission, consider the fact that generating very intense $\Gamma_{em}$ could be difficult in practice. However, caution should be taken as other complexities may appear in practical plasma device. In tokamak for instance, the need to remove helium ash dictates a steady ion flux that shall not be minimized independently. Also there exists strong toroidal magnetic field which dominates over the poloidal field, making an extremely shallow angle at the wall/divertor surface, the emitted electron from the surface would tend to strike back to surface after completing a gyro-orbit. Therefore, more detailed investigations are needed to make use of the ion flux mitigation effect of IRP in practice.

Another promising prospect of IRP is plasma-based material processing. Ion flux is not easy to control in collisionless sheath since it conserves. Invoking IRP can monitor ion flux as well as ion incident energy, according to $\Gamma_{iw} = \Gamma_{i0} \exp(-\frac{e\overline{\varphi_{inv}}}{T_i})$ and Tab I, with $\overline{\varphi_{inv}}$ adjustable with respect to $\Gamma_{em}$ and $V_{RF}$. The maximum ion flux $\Gamma_{i0}$ is instantly available by switching IRP off, offering possibility to control reaction rate in etching, deposition, synthesis, etc[30, 51, 75].

Also, the large RF current in IRP can generate electrostatic waves which may be further converted into electromagnetic waves. Electrostatic waves can be excited by applying RF voltage on matched probe immersed in plasma, to be detected and amplified for measurement[76, 77]. Strong electrostatic waves generated in IRP can be transformed to electromagnetic radiation through mode conversion in inhomogeneous plasma[78, 79], which is expected to be implemented in related experiments.

In the end, instructions will be given to help demonstrate IRP in practice. In some recent works, the existence of inverse sheath in floating condition has been confirmed experimentally. Wang et al made use of electron-induced secondary electron emission.[64]. Secondary electrons are emitted from stainless steel surface, with emission coefficient controlled by primary electron (PE) energy. Stable inverse sheath is formed with high PE energy (over 100 eV) and intense emitted electron density. Alternatively, Kraus and Raitses capitalized on thermionic emission.[52] A thoriated tungsten filament is immersed in plasma. When heated, thermionic current rises up until dominating over plasma electron current, on which a surface potential higher than plasma potential was observed, consistent with inverse sheath theory. In typical condition of RF discharge, thermionic emission is encouraged to trigger IRP since secondary electron emission requires very high electron temperature. A planar 1D configuration with two thermionic emitter seems appropriate.

## 5. Conclusions

In conclusion, we show with simulation and on theoretical ground that boundary emission in RF plasma can produce strong disturbance and establishes a new inverted RF plasma different from normal CCP. Applied bias is mainly consumed by bulk plasma instead of sheath, also external field is not shielded by sheath. It naturally confines ions and shows nonclassical sheath coupling, presheath-sheath structure, particle and energy balance, etc. Invoking inverted RF plasma mitigates ion erosion in plasma boundary where excessive ion flux is damaging. It also provides inspiration for new reaction control technic in plasma processing.

**Acknowledgements**

Guang-Yu Sun would like to thank Dr. M.D. Campanell in Lawrence Livermore National Laboratory (LLNL) for his help in modeling and Dr. I. D. Kaganovich, A.V. Khrabrov in Princeton Plasma Physics Laboratory (PPPL) for fruitful discussions. Authors also thank Prof. Ute Ebert at CWI Amsterdam and Dr. Markus Becker at INP Greifswald for their kind suggestions. This research was conducted under the auspices of National Natural Science Foundation of China (NSFC) under Grant No. 51827809, 51707148, U1830129, U1766218.

**Author contributions**

G.Y. Sun built simulation mode, conducted data analyses, theoretical deductions and wrote the manuscript, A. B. Sun helped build simulation code and theoretical deduction, G. J. Zhang supervised the work throughout. All authors contributed to the compilation and review of the manuscript.